\documentclass[preprint2]{aastex701}
\begin{document}

\title{Doubly Discordant S$H_0$ES NGC4258 Cepheid Relations (HVI), and Impactful Extinction Laws}

\author[orcid=0000-0001-8803-3840]{Daniel Majaess}
\affiliation{Mount Saint Vincent University, Halifax, Nova Scotia, Canada.}
\email[show]{Daniel.Majaess@msvu.ca}

\begin{abstract}
S$H_0$ES 2016-2022 $HVI$ data for classical Cepheids in the keystone galaxy NGC4258 yield doubly discordant Wesenheit Leavitt functions:~$\Delta W_{0,H-VI} = -0.13\pm0^{m}.02$ ($-0^{m}.17$ unweighted) and that is paired with a previously noted $\Delta W_{0,I-VI}\simeq-0^{m}.3$, which in concert with complimentary evidence suggest the 2016 S$H_0$ES NGC4258-anchored $H_0 \pm \sigma_{H_0}$ warrants scrutiny (i.e., $\sigma_{H_0}/{H_0}\gtrsim 6$\%). Cepheid distance uncertainties are further exacerbated by extinction law ambiguities endemic to such Leavitt relations (e.g., NGC4258), particularly for comparatively obscured variables (e.g., $\Delta d \gtrsim 4$\%, reddened Cepheid subsamples in the Milky Way, M31, NGC2442, NGC4424, NGC5643, NGC7250).  Lastly, during the analysis it was identified that the 2022 S$H_0$ES database relays incorrect SMC Cepheid photometry. 
\end{abstract}


\section{Introduction}
\label{sec:intro} 
The keystone galaxy NGC4258 (M106) poses a long-standing challenge to Cepheid researchers, as evidenced by a history of both mismatched distances and Leavitt functions \citep[e.g.,][and references therein]{maj24}. Complications arise partly from degeneracies fostered by notable galactocentric crowding, surface brightness, and chemical abundance trends, and possibly non-standard obscuration parameters. Recently, a substantial $0^{m}.3$ offset was highlighted between the \citet[][S$H_0$ES]{hof16}\footnote{Irrespective of those concerns, \citet[][S$H_0$ES]{hof16} presented a seminally expansive $W_{I-VI}$ sample of Cepheids beyond the Local Group.} and \citet[][S$H_0$ES]{yua22} Wesenheit $W_{I-VI}$ magnitudes for NGC4258 Cepheids  \citep{maj24b}, and inhomogeneities extend to \citet[][]{mao99}, \citet[][]{new01}, \citet[]{mac06}, and \citet{fau15}.  The \citet[][S$H_0$ES]{yua22} findings are favored since they employ an enhanced characterization of complicated photometric contamination (see their \S 4), thereby validating existing concerns calling for such scrutiny of prior S$H_0$ES endeavors \citep[e.g.,][]{maj20}. Importantly, the divergent NGC4258 results are emblematic of broader difficulties inherent to Cepheids \citep[e.g., photometric contamination, standardization, extinction law and metallicity,][see also \S5 of \citealt{mac01}, right panel of Fig.~9 in \citealt{yua20}, \S4 of \citealt{yua22}, and in unison Fig.~1 in \citealt{mf23} and Fig.~6 in \citealt{yua22}]{bon08,maj10,fau15,maj24}. Consequently, TRGB and JAGB distances are desirable, especially when associated with uncrowded and less obscured regions \citep[][]{fm23b}.  Pertinently, \citet[][CCHP]{fre24} concluded that Cepheid and S$H_0$ES distances are too proximate relative to their TRGB and JAGB results, and problems exist prior to the extension by SNe to ascertain $H_0$.  A separate view is espoused by \citet[][S$H_0$ES]{ri24}.

\begin{deluxetable}{ccc}
\tablecaption{NGC4258 S$H_0$ES $2016-2022$ data.\label{table:comp}}
\tablehead{\colhead{S$H_0$ES Data Ref.} & \colhead{$W_{\lambda}$} & \colhead{$\Delta \beta$}}
\startdata
\citealt{rie16} & $H-VI$ & $-0.13\pm0^{m}.02$ \\
\citealt{rie22} & &  \\
\hline
\citealt{hof16} & $I-VI$ & $-0^{m}.3$ \\
\citealt{yua22} & & \\
\enddata
\tablecomments{for $W_{I-VI}$ findings see \citet{maj24b}.}
\end{deluxetable}

\citet[][CCHP]{fre24} relayed JAGB and TRGB expansion rates of 67.96 and 69.85 km s$^{-1}$ Mpc$^{-1}$, while \citet[]{tam13} and \citet[][S$H_0$ES]{ri24} favored $64.1\pm2.0$ and $73.2\pm0.9$ km s$^{-1}$ Mpc$^{-1}$, respectively. Those findings are presented while being ever mindful of the Sandage and de Vaucouleurs $H_0$ debate,\footnote{\citet[][]{ov91}.} which underscores the relevance of blinding procedures \citep[e.g., \S 4 in][CCHP]{fre24}. On that broader topic \citet[][DESI]{ch24} advocate, \textit{`To avoid confirmation bias ...~blinding procedures become a standard practice in the cosmological analyses of such surveys.'} Yet S$H_0$ES lacks a published comprehensive blinding process, and the \citet{rie05,ri24} $H_0$ remained comparatively unaltered across nearly two decades despite internal inconsistencies, which challenge the notion of robustness and cumulative convergence. Research by \citet{efs20} implies the following: vetting S$H_0$ES is partly hindered by unpublished pre-culled Cepheid datasets, there are Leavitt slope offsets \citep[e.g., see also Fig.~2 in][]{maj10}, the uncertainties are underestimated \citep[e.g., see also Table~1 in][]{maj24b}, and sizable color offsets exist \citep[e.g., see also bottom left panel of Fig.~6 in][]{maj10}. Certain S$H_0$ES compilations can be challenging to scrutinize because their $VI$ photometry isn't decoupled into fully decontaminated separate bands, and early S$H_0$ES data should be interpreted cautiously since the $W_{V-VI}$ slope \citet[][S$H_0$ES]{rie09a} determined for solar-Cepheids contradicted Local Group counterparts \citep[compare Fig.~12 in \citealt{rie09a} to Fig.~2 in][and Fig.~1 in \citealt{maj11}]{maj10}. The shallower slope determined by \citet[][S$H_0$ES]{rie09a} may be indicative of non-standard photometry or inaccurate decontamination \citep[e.g., Fig.~2 in][]{maj20}, and furthermore a subset of S$H_0$ES observations were too blue \citep[e.g., NGC1309,][]{maj10}.  Moreover, cross-referencing S$H_0$ES Cepheids solely on the basis of coordinates can prove unsatisfactory \citep[see also][and discussion therein]{efs20}.  Issues expressed regarding S$H_0$ES are conveyed in several studies \citep[e.g.,][see also \citealt{maj10,maj24,maj24b}]{efs20,mor22,mf23,wo24,fre24,bl24}. 

Concurrently, rational foundational concerns persist relative to $\Lambda$CDM \citep[e.g.,][]{st11,kr12,lc14,pe15},\footnote{Significant issues associated with canonical cosmology include:~\textit{`...~dark matter particle or not, explanation of cosmic acceleration, the transformation of inflation into a fundamental theory.'} - UChicago Cosmic Controversies conference.} and hence $H_0$ tied to that model \citep[e.g., see \S 1 and Appendix A in][]{rb23}. A summary of earlier-epoch $H_0$ $\Lambda$CDM estimates by the NASA LAMBDA team included:~SPTPol 2017 ($71.2\pm2.1$ km s$^{-1}$ Mpc$^{-1}$), Planck PR3 2018 ($67.36\pm0.54$ km s$^{-1}$ Mpc$^{-1}$), Planck$+$ACTPol$+$SPTPolEE 2021 ($68.7\pm1.3$ km s$^{-1}$ Mpc$^{-1}$).  The spread partly arises from inhomogeneous coverage, yet expands when considering models beyond canonical $\Lambda$CDM. That extended ensuing baseline can be compared to relatively local independent $H_0$ determinations \citep{ste20}.  

In this study, additional issues regarding NGC4258 data are relayed.  In \S \ref{sec:analysis} $W_{H-VI}$ magnitudes tied to S$H_0$ES 2016$-$2022 NGC4258 observations are compared.  In \S \ref{sec:ext} the compounding impact of an uncertain ratio of total-to-selective extinction $R$ on such Wesenheit magnitudes is reassessed, and systematic distance shifts are characterized. $R$ linked to NGC4258 may be anomalous \citep[][and caveats therein]{fau15}, and a debate continues regarding the broader ramifications of $R$-uncertainties \citep[][and see \citealt{rie22} for the counterpoint]{mor22,wo24}. 

\section{Analysis}
\subsection{Photometric inhomogeneities}
\label{sec:analysis}
A Wesenheit formulated Leavitt Law is:
\begin{eqnarray}
\label{eqn:beta}
W_{H-VI}&-&W_{0,H-VI} = \mu_0  \\ 
(H-R_{H-VI}(V-I)) &-& (\alpha \log{P} + \beta) = \mu_0 \nonumber 
\end{eqnarray}
For example the filters can be H (F160W), V (F555W), and I (F814W). The selected ratio of total-to-selective extinction ($R_{H-VI}\approx0.4$) is the midpoint cited by \citet[][S$H_0$ES]{rie16}, who noted, \textit{`ranging from 0.3 to 0.5 at H depending on the reddening law'} \citep[see also Table~6 in][]{rie22}. The impact of that spread is discussed in \S \ref{sec:ext}.  The slope ($\alpha$) stems from an analysis of \citet[][S$H_0$ES]{rie19} observations for LMC Cepheids, owing to the relative insensitivity of that term to metallicity in the passbands examined \citep[e.g.,][]{rie22}. 

A broader problem is promptly elucidated by comparing zeropoints of the absolute Wesenheit magnitude relation inferred from \citet[][]{rie19} S$H_0$ES LMC Cepheids ($\beta=-2.53\pm0^{m}.03$), \citet[][]{rie16} S$H_0$ES NGC4258 Cepheids ($\beta=-2.72\pm0^{m}.04$), and \citet[][]{rie22} S$H_0$ES NGC4258 Cepheids  ($\beta=-2.59\pm0^{m}.03$).  Those results are tied to the \citet{pie19} LMC and \citet{rei19} NGC4258 anchor points, following S$H_0$ES.  In concert with the \citet{maj24b} findings (their \S 2), the results point to problems endemic to the 2016 S$H_0$ES $VIH$ NGC4258-based $H_0$ and $\sigma_{H_0}$ \citep[Tables~6 and 8 in][]{rie16}.\footnote{See also the important findings conveyed in \S 10 of \citet{fm23b}.}  Alternatively, when the same galaxy is utilized its distance (\& uncertainty) can be obviated and the significance of the offset strengthens, whereby NGC4258 results inferred from \citet[][S$H_0$ES]{rie16,rie22} subtract to yield:~$\Delta \beta =  -0.13\pm0^{m}.02$ or an unweighted $-0^{m}.17 \pm 0.03$ (Table~\ref{table:comp}).  The problems arise owing to an apparent Wesenheit magnitude offset between those NGC4258 datasets ($\Delta W_{H-VI}$), since there are $\Delta H$$\Delta (V-I)$ photometric deviations amongst common stars.  That is the cause, and the effect can be observed upon $\beta$ or $W_{0,H-VI}$.  That is likewise true of the \citet[][S$H_0$ES]{hof16} and \citet[][S$H_0$ES]{yua22} NGC4258 discrepancy (see also \S 4 of the latter).  A \textit{contested} metallicity effect does not reconcile all aforementioned data \citep[for contrasting opinions on the metallicity topic see][]{mf23,rb23}.

\subsection{$R$-ambiguities}
\label{sec:ext}
\citet[]{fau15} posited that NGC4258 could adhere to an anomalous $R_V\simeq 4.9$ relative to the Milky Way \citep[$R_V\simeq 3.26$,][]{ber96}, yet cautioned and preferred that, \textit{`it seems probable that at least one other systematic effect is at work in our sample's colours.'} Furthermore, a galaxy could possess $R$-variations as a function of galactocentric distance \citep[e.g.,][]{gon13}, in tandem with a mean diverging from a broader extragalactic sample \citep[e.g.,][]{go03}.  \citet{fi99} remarked that a \textit{`bewildering variety of IR-through-UV extinction curves'} are commonplace, and \citet[][]{go03} noted that Milky Way and SMC extinction curves exhibit a \textit{`continuum of properties'} \citep[see also Tables~2 and 4 in][and Table~1 in \citealt{tur13}]{tu76}. Specifically, the Galactic Bulge \textit{may} follow a separate $R_{VI}$ relative to the Solar Neighborhood \citep[e.g.,][]{uda03}, and \citet{maj16} confirmed that Carina features an extreme optical $R_{V}$ \citep[e.g.,][]{tur13,car13}, whereas the near-infrared $R_{J-JK_s}$ results were comparatively constant across $\ell$.  A debate persists whether infrared extinction laws vary \citep[e.g.,][]{zas09}, and if a subset of passbands within that domain are acutely sensitive to compositional changes \citep[e.g.,][]{hg74,sc11,maj13e}. 

As noted $R_{H-VI}$ could span $\simeq 0.3 - 0.5$, while $R_{I-VI}$ may traverse $\simeq 1.30 - 1.55$, and those estimates are conservative \citep[e.g., Table~2 in][and Table~6 in \citealt{rie22}]{uda03}.  For example \citet[][S$H_0$ES]{hof16} adopted $R_{I-VI}=1.45$, whereas \citet[][S$H_0$ES]{yua22} selected $R_{I-VI}=1.30$ following \citet[][S$H_0$ES, and references therein]{rie19}.  Regarding the $W_{H-VI}$ passband combination, \citet[][S$H_0$ES]{rie09a} favored $R_{H-VI}=0.479$, whereas \citet[][S$H_0$ES]{rie22} derived $R_{H-VI}=0.34$ from their full Cepheid sample. 

In sum, there exists a breadth in the potential extinction law employed for extragalactic Cepheids.  \textit{An avenue} to approximate the impact on distances by the aforementioned spread is to derive absolute and apparent Wesenheit functions using the same $R$. First, the \citet[][S$H_0$ES]{rie19} LMC photometry was utilized to establish the absolute Wesenheit magnitude, in tandem with the Araucaria anchor.\footnote{Such anchors may require revision (importantly, possibly unidirectionally), and separate determinations exist that possess low \textit{cited} uncertainties \citep[][and references therein]{ste20}.}  Second, distances were subsequently computed for individual Galactic Cepheids by relying on \citet[][S$H_0$ES]{rie21} for the apparent Wesenheit computation.  That S$H_0$ES Milky Way photometry advantageously samples a sizable extinction baseline.  The ensuing median distance offset between $R_{I-VI}$ extrema is $\simeq 5$\%, with lower $R_{I-VI}$ yielding further distances.  For $R_{H-VI}$ a $\simeq 4$\% median difference was identified.  The largest distances established using $R_{H-VI}=0.30$, relative to the shortest determined via $R_{I-VI}=1.55$, were separated by $\simeq 8$\% (median). $R_{H-VI}$ extrema distances were generally more remote than those tied to $R_{I-VI}$.

The broadly illustrative approach was likewise applied to the expansive extragalactic data of \citet[][S$H_0$ES]{rie16,rie22} and \citet[][S$H_0$ES]{hof16}, but this time for all $P>7^{d}$ Cepheids within a given galaxy (i.e., period criterion mitigates overtone contaminants).  Cepheids in M31, NGC2442, NGC4424, NGC5643, and NGC7250 featured sightlines with comparatively enhanced obscuration, and possessed a $\gtrsim 4$\% distance offset.  NGC4258 falls short of that group ($\Delta d \lesssim1.6$\%, dataset depending) if $R$ lies within the aforementioned bounds and is not anomalous \citep[however see][and caveats therein]{fau15}. That uncertainty is paired with the more dominant $W_{H-VI}$ S$H_0$ES photometric offset (\S \ref{sec:analysis}), which implies\footnote{Eqn.~2.1 in \citet[][]{efs20}.} a $\approx 6$\% $H_0$ internal discrepancy.  The $H_0$ inconsistency is significantly higher amongst the $W_{I-VI}$ S$H_0$ES 2016-2022 data \citep[][]{maj24b}. 

Lastly, the macro data inspection indicated that SMC Cepheid $W_{H-VI}$ photometry compiled by \citet[][S$H_0$ES]{rie22} appears awry, and overlays upon the LMC Wesenheit function.  That conclusion was confirmed by constructing a sample of SMC Cepheids by drawing from photometric catalogs in the literature. More broadly, the SMC can pose a comparatively greater challenge as a Wesenheit calibrator (e.g., LMC) granted its sizable depth along the sightline and inclination \citep[][e.g., the latter's Fig.~12]{cc86,gr00,ss12}.

\section{Conclusions}
The absolute $W_{0,H-VI}$ magnitude zeropoint for Cepheids in the critical maser galaxy NGC4258 spans $>0^{m}.1$ across the S$H_0$ES $2016-2022$ datasets (Table~\ref{table:comp}). That adds to a $0^{m}.3$ $W_{0,I-VI}$ discrepancy unveiled previously and linked to a separate passband combination \citep{maj24b}, and points to problematic 2016 S$H_0$ES NGC4258 photometry and $H_0\pm \sigma_{H_0}$ anchored to that galaxy \citep[see entries in Tables~6 and 8 in][]{rie16}.   Moreover, ensuing Cepheid distance uncertainties are further enlarged beyond that S$H_0$ES internal discrepancy (an implied $\gtrsim 6$\% $H_0$ when associated solely with that maser anchor) for relatively reddened subsamples owing to $R_{H-VI}$ and $R_{I-VI}$ ambiguities (\S \ref{sec:ext}).  

Continued independent research on the broader impact of $R$-variations on Cepheid \textit{and} SN distances is desirable \citep[e.g.,][]{er06,go08,fau15,wo24}, as exemplified by the ambiguity regarding the distance to Centaurus A \citep{fe07,maj10}. Constraints could emerge from breakthroughs in characterizing the interstellar medium by revealing source(s) behind the 220 nm extinction bump, numerous unidentified infrared emission lines, and $> 500$ diffuse interstellar bands \citep[e.g.,][]{tur14,xi17,eb24,maj25b}.  

The S$H_0$ES team conveys their perspective on how $R$-uncertainties affect $H_0$ in several studies \citep[e.g.,][and the latter's \S 6.3 and Appendix D]{rie09a,rie22}, and for separate opinions see \citet{mor22} and \citet{wo24}.

\bibliography{article}{}
\bibliographystyle{aasjournalv7}

\end{document}